\documentclass[reprint, superscriptaddress, onecolumn, aps, pre]{revtex4-2}
\usepackage{amsfonts, amssymb, amsmath}
\usepackage{epstopdf}
\usepackage{graphicx}
\usepackage{epsfig}
\usepackage{hyperref}
\usepackage{color}

\newcommand{\zz}{{\bf \hat z}}

\begin{document}
\title{Four-dimensional equations for the study of electromagnetic plasma turbulence in a drift kinetic limit}
\author{Evgeny A. Gorbunov}
\affiliation{Coventry University, Coventry CV1 5FB, United Kingdom}
\author{Bogdan Teaca}
\affiliation{Coventry University, Coventry CV1 5FB, United Kingdom}
\affiliation{University of Craiova, 13 A.I. Cuza Street, 200585 Craiova, Romania}
\begin{abstract}
For a magnetised plasma in a straight magnetic guide field, we derive a set of four-dimensional kinetic equations, which can capture electromagnetic turbulence in the drift kinetic limit. To do so, we start from the gyrokinetic equations, employ a Laguerre decomposition in the perpendicular velocity direction, retain only the dominant gyroaverage contributions and only the first two Laguerre moments that source the electromagnetic fluctuations. The model conserves free energy, and can describe electromagnetic turbulence for a plasma at the transition between fluid and gyrokinetic regimes ($k_\perp \rho_i\approx 1$ range of scales), as dominant finite-Larmor-radius (FLR) effects are considered. In addition to the three dimensions in positions space, we retain the parallel velocity dependence, which we describe via a Hermite representation. Employing this system, but without any other physics based assumptions for the plasma species that can bias results, will allow us to investigate how fluid effects transition into the kinetic range, and analyse the interplay between spatial and velocity space mixing for electromagnetic plasma turbulence. 
\end{abstract}
\maketitle

\section{Introduction}

In astrophysical settings, the use of magnetohydrodynamics (MHD) is successful in the description of the large scale dynamics of plasma turbulence \cite[e.g.][]{Goldreich:1995p724, Bhattacharjee:1998p313, Zhou:2004p21, Chandran:2009p760, Zhdankin:2012p1824, Weidl:2015p1676}. At the same time, the use of kinetic formalisms allows us to probe micro-scales in the collisionless limit \cite[e.g.][]{Wan:2012p1837, Servidio:2012p1845, Cerri:2018p2139, 2019PhRvX...9c1037G}. Gyrokinetic (GK) theory in particular is used for the study of turbulence in collisionless magnetised plasma at sub-gyroradius scales \cite{Howes:2006p1280, Howes:2008p1132, Howes:2011p1370, Tatsuno:2009p1096, Tenbarge:2013p1730, Told:2015p1712, Navarro:2016p1965, Teaca:2019p2154, Teaca:SGS2021}. Theoretical considerations \cite{Schekochihin:2008p1034, Schekochihin_2009}, numerical simulations \cite{Tatsuno:2009p1096, Told:2015p1712} and satellite observations \cite{Alexandrova:2009p1396, Alexandrova:2008p1397} have shown that at the ion Larmor (gyroradius) scale, as turbulence transitions from a fluid to a kinetic dynamical regime, a break occurs in the spectra of various quantities. While the scientific community is aware of this spectral break for the last decade, recent observations from the Parker Solar Probe \cite[e.g.][]{Zhao_2020,Vech_2020,Duan_2020, Duan_2021} have spurred an interest for understanding the dynamics of this transition. 

To study this transition in detail, while accounting for the collisionless regime, a GK formalism can be used for magnetised plasma. Even when the strength of the magnetic guide fields is not that large and the GK ordering is weakened, such as in the solar wind, GK theory is still useful in capturing the main dynamical properties, such as the kinetic Alfv\'en waves (KAW) cascade, Landau damping, linear and nonlinear phase space mixing. While the five-dimensional phase space for GK represents a simplification over the full Maxwell-Vlasov system, it is still too cumbersome to use for systematic studies of the fluid-kinetic transition. For this reason, starting from the electromagnetic GK equations, we will derive a four-dimensional drift kinetic system that accounts for the dominant finite Larmor radius (FLR) effects. We preserve the three spatial directions and the parallel velocity direction that will allow us to account for the important parallel phase mixing dynamics.   

In essence, starting from the GK equations and employing a Laguerre decomposition in the perpendicular velocity direction (\S\ref{section:derivation}), retaining only the dominant gyroaverage contributions and only the first two Laguerre moments for the distribution functions, we obtain a closed set of four-dimensional kinetic equations valid in the large-scale drift kinetic limit (\S\ref{section:Drift-kinetic limit}). In addition to the three-dimensions in positions space, we retain the parallel velocity direction, which we describe via a Hermite representation (\S\ref{section:spectraleq}). The model we will propose conserves free energy and describes electromagnetic turbulence for a plasma at the transition between fluid and gyrokinetic regimes ($k_\perp \rho_i\approx 1$ range of scales), accounting for FLR effects. It allows for Landau damping, and can allow or potentially prohibit parallel mixing to occur due to turbulent plasma echos \cite{schekochihin_2016,Kawazura771,Meyrand2019}. The strength of the model consists in linking with GK at $k_\perp \rho_i> 1$ scales, while allowing for any (gyro) fluid nature for the ions or electrons species to emerge naturally and without any other physical assumption. Numerically, it will represent a good compromise that will allow for the study of how fluid constraints (similar to magnetic-helicity and cross-helicity in MHD) can impact the transition towards the kinetic dynamics, or how the microphysics affect the macrophysics of the plasma \cite{meyrand2021}. Considering that the nonlinear interaction of spatial structures and the velocity phase mixing in the parallel direction represent the main dynamics of kinetic turbulence, the model also offers a simplified system for the study of the interplay between spatial turbulence and parallel phase mixing when electromagnetic interactions are present.

\section{GK equations in Laguerre space as a starting point}\label{section:derivation}

\subsection{The gyrokinetic equations}\label{section:GKeq}

The starting point for our derivation is given by the gyrokinetic (GK) equations in the presence of a straight magnetic guide field (acting here in the $z$-direction, $B_0\zz$). The community regularly employs the GK equations in the study of kinetic plasma at sub-gyroradius scales ($k_\perp \rho_i > 1$), and we have used them to probe the fundamental nature of GK turbulence \cite{Teaca:2012p1415,Teaca:2014p1571,Teaca:2017p1989,Teaca:2019p2154,Teaca:SGS2021}. A simple $\delta f$-derivation of the GK equations for a Maxwellian background distribution function ($F_s$) is presented in \cite{Howes:2006p1280, Schekochihin_2009} and as such, we will not insist here on the GK construction.

For $h_s = h_s(\mathbf{R}_s,v_{\parallel},\mu)$ the non-adiabatic part of the gyrokinetic distribution function (with the particle perturbed distribution function being found as $\delta f_s= - q_s \phi F_s/T_s + h_s$), we simply list below the GK equations in non-dimensional form and in the absence of collisions
\begin{equation} \label{eq:Vlasov}
    \frac{\partial h_s}{\partial t}+ \frac{1}{B_0}\{\chi_s,  h_s\} + v_{T_s}v_{\parallel}\frac{\partial h_s}{\partial z}=\frac{q_s }{T_{s}}F_s\frac{\partial \chi_s}{\partial t}\,,
\end{equation}
where $\{\chi_s,  h_s\}=(\nabla \chi_s \times \nabla h_s)\cdot {\bf \hat z}$ is the nonlinearity written in term of a Poisson bracket structure, $v_{Ts}=(\sqrt{2T_s/m_s})$ is the thermal velocity for a species $s$ of charge $q_s$, mass $m_s$ and background temperature $T_s$ (all expressed in terms of reference units). The perpendicular velocity dependence is expressed in term of the magnetic moment $\mu$ (in non-dimensional form, $\mu B_0= v_\perp^2 $) and the equilibrium distribution function $F_s$ has the simple form
\begin{align}
F_s = \pi^{-3/2}e^{-{v_\parallel^2}-\mu B_0} \,.
\end{align}

Here, $\chi_s$ is the gyroaveraged gyrokinetic potential. It is convenient to represent it in Fourier space, as the gyroaverage operations reduce to the multiplication of Bessel functions of zero ($J_0$) and first ($J_1$) order,
\begin{equation} \label{eq:chi_k}
\chi_s(\mathbf{k},\mu,v_\parallel) = \sum_\mathbf{k} e^{i\mathbf{k}\cdot\mathbf{R}_s}\left[J_0\left(\lambda_s\right)\left(\phi(\mathbf{k})-v_{T_s} v_\parallel  A_\parallel(\mathbf{k})\right)+\frac{\mu T_{s}}{q_s}\frac{2 J_1\left(\lambda_s\right)}{\lambda_s}B_\parallel(\mathbf{k})\right]\,.
\end{equation}
The electrostatic potential ($\phi$), magnetic potential in the parallel direction ($A_\parallel$), and magnetic fluctuation in the parallel direction ($B_{\parallel}$) are obtained in wave space from their respective field equations,
\begin{align}
\phi(\mathbf{k},t) &= \pi B_0\sum_s  q_s n_s \int^{+\infty}_{-\infty}dv_\parallel \int_0^{\infty}d\mu J_0(\lambda_s)h_s(\mathbf{k},v_\parallel,\mu,t)\bigg{/} \sum_s\frac{q_s^2 n_s}{T_s}\,,\label{eq:phi} \\
A_\parallel(\mathbf{k},t) &= \frac{\pi \beta}{2 k_\perp^2}B_0\sum_s q_s n_s v_{T_s} \int^{+\infty}_{-\infty}dv_\parallel v_\parallel \int_0^{\infty}d\mu  J_0\left(\lambda_s\right)h_s(\mathbf{k},v_\parallel,\mu,t)\,,\label{eq:A}\\
B_\parallel(\mathbf{k},t) &= -\frac{\pi\beta}{4}B_0\sum_s m_s n_s v_{T_s}^2  \int^{+\infty}_{-\infty}dv_\parallel \int_0^{\infty}d\mu\ \mu \frac{2 J_1\left(\lambda_s\right)}{\lambda_s}h_s(\mathbf{k},v_\parallel,\mu,t)\,, \label{eq:B}
\end{align}
where $\beta=8\pi n_{\mbox{{\scriptsize ref}}} T_{\mbox{{\scriptsize ref}}}/B_{\mbox{{\scriptsize ref}}}^{2}$ is defined in term of the reference temperature $T_{\mbox{{\scriptsize ref}}}$, density $n_{\mbox{{\scriptsize ref}}}$ and magnetic intensity $B_{\mbox{{\scriptsize ref}}}$. We also use
\begin{align}
    \lambda_s &= \sqrt{2\mu B_0 b_s}\,, \label{eq:lambda}\\
    b_s &= \left(\frac{k_\perp v_{Ts}}{\sqrt{2}\Omega_s}\right)^2 =\frac{k^2_\perp \rho_s^2}{2} \,. \label{eq:b} 
\end{align}

From the GK equations above, we see that the gyroaverage operations, captured by the Bessel functions, smear $k_\perp$ structures into $\mu$ structures at $k_\perp \rho_i > 1$. In the $k_\perp \rho_i \ll1$ limit, for which the Bessel functions tend towards one, if no initial $\mu$ structure is present in $h_s$, the GK equations cannot generate subsequent $\mu$ structures and we can integrate over the $\mu$ direction without any loss of information. To see this better, and to integrate $\mu$ in a consistent way, we make use of a Laguerre transform approach.

\subsection{Laguerre transform} \label{section:Laguerre transform}

A recent detailed approach for the use of a Laguerre transform for the GK system can be found in \cite{mandell2018}. Here, we present only the information needed to understand the transformations we will employ in our derivation. A notable simplification is that we resume to a constant $B_0$ rather than the more general $B_0(z)$ approach employed by \cite{mandell2018}. However, we do not resume to the simpler electrostatic limit, and consider electromagnetic fluctuations in our derivation.  

For convenience, we define the Laguerre (basis) functions as
\begin{align}
\psi^l\left(\mu B_0\right) &= (-1)^l e^{-\mu B_0} L_l\left(\mu B_0\right)\,,\\
\psi_l\left(\mu B_0\right) &= (-1)^l L_l\left(\mu B_0\right) \,,
\end{align}
where $L_l(x) = \frac{e^x}{l!}\frac{d^l}{dx^l}x^l e^{-x}$ are Laguerre polynomials. The Laguerre functions obey the orthonormality condition,
\begin{equation}\label{orth}
    \int_{0}^{+\infty}d\mu B_0 \psi^k\psi_l = \delta_{kl}\,,
\end{equation}
which results from the orthogonality condition of the Laguerre polynomials $L_l(x)$ around the $e^{-x}$ weight function (note that $\psi_l\left(x\right) e^{-x}=\psi^l\left(x\right)$). In our derivation, we will also make use of the recurrence relation
\begin{align}\label{recurrence}
    (l+1)\psi^{l+1} = (\mu B_0-2l-1)\psi^l - l\psi^{l-1}\,.
\end{align}

The decomposition of a function $f(\mu)$ in term of the Laguerre basis $\psi^l$ is done as
\begin{align}\label{equation:Laguerre basis}
f(\mu) = \sum_{l=0}^{\infty} \psi^l\left(\mu B_0\right) \widehat{f}_l\,,
\end{align}
with the spectral amplitude $\widehat{f}_l$ for the Laguerre modes $l$ being found simply as
\begin{align}\label{equation:spectral amplitude}
\widehat{f}_l = \int_0^{\infty}d\mu B_0 \psi_l f(\mu)\,.
\end{align}
 
The Laguerre transforms given by eqs. \ref{equation:Laguerre basis}-\ref{equation:spectral amplitude} allows us to obtain the GK equations in Laguerre space. Apart from projecting $h_s$, we will also need to project $J_0(\lambda_s)$ and $\frac{2}{\lambda_s} J_1(\lambda_s)$ onto the Laguerre basis. Note that $\chi_s$ depends on $\mu$ solely through the Bessel functions. In the drift kinetic limit, retaining the contribution of the Bessel functions will allow for finite Larmor radius (FLR) effects to be preserved \cite{Dorland93}. 

Remembering that $\lambda_s = \sqrt{2\mu B_0 b_s}$, and omitting the species  subscript $s$, we write the Bessel functions of order $n$ in term of the basis $\psi_l$, with the coefficients $\mathcal{J}_{nl}$ obtained via a generating function, i.e. 
\begin{align}\label{decompositionBessel}
   J_n\left(\sqrt{2\mu B_0 b }\right)  = \sum_l \psi_l \mathcal{J}_{nl} = \sum_l \psi_l \frac{b^l}{l!}\frac{\partial^l}{\partial b^l} \langle J_n \rangle\,,
\end{align}
where
\begin{align}\label{equation:Bessel zero moment}
   \langle J_n \rangle=\int_0^{\infty} e^{-\mu B_0} J_n\left(\sqrt{2\mu B_0 b}\right) B_0 d\mu\,.
\end{align}

Obtaining the same decomposition for the $\frac{2}{\lambda_s} J_1(\lambda_s)$ function requires an extra manipulation. Using the well-known recurrence formula for the Bessel functions, we arrive at the expression
\begin{equation}\label{equation:Bessel first order}
   \frac{2}{\lambda_s} J_1\left(\lambda_s\right)  = J_0\left(\lambda_s\right)+J_2\left(\lambda_s\right) = \sum_l \psi_l \mathcal{\tilde{J}}_{1l}=\sum_l \psi_l \frac{b^l}{l!}\frac{\partial^l}{\partial b^l} (\langle J_0 \rangle+\langle J_2 \rangle)\,.
\end{equation}
The coefficients for the $\frac{2}{\lambda_s} J_1\left(\lambda_s\right)$ function are labeled as $\mathcal{\tilde{J}}_{1l}$ (the tilde notation is used to distinguish it form the coefficients of $J_1\left(\lambda_s\right)$) and we express it via the generating function involving $\langle J_0 \rangle$ and $\langle J_2 \rangle$. From \eqref{equation:Bessel zero moment} we obtain 
\begin{align}
\langle J_0 \rangle &= e^{-b/2}\,,\\
\langle J_2 \rangle &= -\left(\frac{2}{b}+1\right)e^{-b/2} + \frac{2}{b}\,,
\end{align}
and the coefficients for the Bessel functions of interest are obtained as
\begin{align}
\mathcal{J}_{0l} &= \frac{b^l}{l!}\frac{\partial^l}{\partial b^l} e^{-b/2} \label{equation:J round 0}\,,\\
\mathcal{\tilde{J}}_{1l} &=  \frac{b^l}{l!}\frac{\partial^l}{\partial b^l} \left[\left(1-e^{-b/2}\right)\frac{2}{b}\right]\,.\label{equation:J round 1}
\end{align}
Compared to \cite{mandell2018}, since we account for electromagnetic fluctuations, we needed to obtain $\mathcal{\tilde{J}}_{1l}$. For clarity, the GK equation in Laguerre space are presented next.

\subsection{Gyrokinetic system in Laguerre basis}

We start by looking at the field equations in Laguerre basis. For $\phi$, given by eq.~\eqref{eq:phi}, we use \eqref{equation:Laguerre basis} for the representation of $h_s$ and \eqref{decompositionBessel} for the representation of $J_0(\lambda_s)$. We obtain
\begin{align}
        \phi(\mathbf{k},t) &= \pi B_0\sum_s  q_s n_s\int^{+\infty}_{-\infty}dv_\parallel \int_0^{\infty}d\mu \sum_l \psi_l\mathcal{J}_{0l}\sum_m \psi^m \widehat{h}_{sm}\bigg{/}\sum_s\frac{q_s^2 n_s}{T_s} \nonumber \\ 
        & =\pi\sum_s  q_s n_s\int^{+\infty}_{-\infty}dv_\parallel \sum_l \mathcal{J}_{0l} \widehat{h}_{sl} \bigg{/} \sum_s\frac{q_s^2 n_s}{T_s}\,,  \label{phil}
\end{align}
where we used the orthogonality condition \eqref{orth} to obtain the second line. The same approach is used for $A_\parallel$, which gives
\begin{equation} \label{Al}
    A_\parallel(\mathbf{k},t) = \frac{\pi \beta}{2 k_\perp^2}\sum_s  q_s n_s v_{T_s}\int^{+\infty}_{-\infty}dv_\parallel v_\parallel \sum_l \mathcal{J}_{0l} \widehat{h}_{sl} \,.   
\end{equation}
For the $B_\parallel$ field equation, after the Laguerre representation for each contributing function is inserted in the first step, we extract $\mu \psi^m$ from the recurrence relation \eqref{recurrence} in the second step, and use the orthogonality condition \eqref{orth} in the last step, obtaining
\begin{align}
B_\parallel(\mathbf{k},t) &= -\frac{\pi\beta}{4}B_0\sum_s m_s n_s v_{T_s}^2 \int^{+\infty}_{-\infty}dv_\parallel \int_0^{\infty}d\mu \mu \sum_l \psi_l \mathcal{\tilde J}_{1l}\sum_m \psi^m\widehat h_{sm} \nonumber\\
 &= -\frac{\pi\beta}{4}\sum_s m_s n_s v_{T_s}^2\int^{+\infty}_{-\infty}dv_\parallel \int_0^{\infty}d\mu \sum_l\psi_l \mathcal{J}_{1,l}\sum_m \big{[}(2m+1)\psi^m +\nonumber \\ 
&\hspace{65mm} m\psi^{m-1}+(m+1)\psi^{m-1} \big{]}\widehat{h}_{sm} \nonumber\\
& = -\frac{\pi\beta}{2}\sum_s \frac{n_s T_s}{B_0}\int^{+\infty}_{-\infty}dv_\parallel 
\sum_m \mathcal{J}_{1,l}  \big{[}(2l+1)\widehat{h}_{s,l} +l \widehat{h}_{s,l-1}+(l+1)\widehat{h}_{s,l+1} \label{Bl}  \big{]}\,.
\end{align}

With the electromagnetic potentials and fields determined, to project the gyrokinetic potential \eqref{eq:chi_k} onto the Laguerre basis, we need to express the Bessel functions contributions and use again the recurrence relation \eqref{recurrence} to express $\mu \psi_m$, before employing the orthogonality condition \eqref{orth}. The $\widehat{\chi}_l$ spectral amplitude gives
\begin{align}
  \widehat{\chi}_l(\mathbf{k}) 
  &= \int_0^{\infty}d\mu B_0 \psi^l
  \left[\left(\phi(\mathbf{k})-v_{T_s} v_\parallel  A_\parallel(\mathbf{k})\right)\sum_m \psi_m J_{0l}+\frac{\mu T_{s}}{q_s}B_\parallel(\mathbf{k})\sum_m \psi_m \tilde{\mathcal{J}_{1l}}\right] \nonumber\\
   &=  B_0\int_0^{\infty} d \mu \psi^l\left(\phi(\mathbf{k})-v_{T_s} v_\parallel  A_\parallel(\mathbf{k})\right)\sum_m \psi_m J_{0m}+\nonumber\\
      &\hspace{20mm}+ \frac{T_s}{q_s B_0} B_\parallel(\mathbf{k})\sum_m\Big[(m+1)\psi_{m+1}+(2m+1)\psi_{m}+m\psi_{m-1}\Big] \tilde{\mathcal{J}}_{1m}
  \mathcal{\tilde{J}}_{1l} \nonumber\\
&= \mathcal{J}_{0l}\left(\phi(\mathbf{k})-v_{T_s} v_\parallel  A_\parallel(\mathbf{k})\right) + \frac{T_s}{q_s B_0} B_\parallel(\mathbf{k})\left((l+1)\tilde{\mathcal{J}}_{1,l+1}+
   (2l+1)\tilde{\mathcal{J}_{1,l}}
   +l\tilde{\mathcal{J}}_{1,l-1}\right)\label{chil}\,.
\end{align}

To obtain the GK equation \eqref{eq:Vlasov} on the Laguerre basis, we look at each term independently. We remind the reader that the $\mu$ dependence of the potential $\chi$ is due to the Bessel functions, and as such, the Laguerre expansion of $\chi$ occurs in term of the basis $\psi_l$. The linear terms are computed in a straightforward maner,
\begin{align}
&\int_0^{\infty} d \mu B_0 \psi_l \sum_m \psi^m \frac{\partial \widehat{h}_m}{\partial t} = \frac{\partial \widehat{h}_l}{\partial t}\,,\\
&\int_0^{\infty} d \mu B_0 \psi_l \sum_m \psi^m v_{T_s}v_{\parallel}\frac{\partial \widehat{h}_m}{\partial z} = v_{T_s}v_{\parallel}\frac{\partial \widehat{h}_l}{\partial z}\,,\\
&\int_0^{\infty} d \mu B_0 \frac{q_s}{T_s} \pi^{-3/2} e^{-v^2_{\parallel} - \mu B_0} \psi_l \sum_n \psi_n \frac{\partial \widehat{\chi}_n}{\partial t} = \frac{q_s}{T_s}  \frac{ e^{-v^2_{\parallel}}}{\pi^{3/2}}  \frac{\partial \widehat{\chi}_l}{\partial t}\,.
\end{align}
The only complication arises from the nonlinear term, which, as presented in \cite{mandell2018}, is expressed as a convolution
\begin{equation}
\int_0^{\infty} d \mu B_0 \frac{1}{B_0}\psi_l \left\{\sum_n \psi_n \widehat{\chi}_n, \sum_\kappa \psi_\kappa \widehat{h}_\kappa\right\} = \frac{1}{B_0}\sum_{\kappa = 0}^\infty\sum_{n = |\kappa-l|}^{\kappa+l} \alpha_{l\kappa n}
\left\{ \widehat{\chi}_n,  \widehat{h}_\kappa\right\}\,,
\end{equation}
where $\alpha_{l\kappa n}$ is a convolution coefficient computed as
\begin{equation}\label{nonlinearl}
    \alpha_{l\kappa n} = \int_0^{\infty} d\mu B_0 \psi_\kappa \psi_n\psi^l = \sum_j \frac{(\kappa+l-j)! 2^{2j-\kappa-l+n}}{(\kappa-j)!(l-j)!(2j-\kappa-l+n)!(\kappa+l-n-j)!}\,,
\end{equation}
with the summation limits over $j$ being chosen to ensure that the factorials under the sum remain non-negative. Putting all the terms together, gives the GK  equation for a Laguerre mode $l$ as
\begin{equation}\label{Vlasovl}
     \frac{\partial \widehat{h}_l}{\partial t} + \frac{1}{B_0}\sum_{\kappa = 0}^\infty\sum_{n = |\kappa-l|}^{\kappa+l} \alpha_{l\kappa n}
\left\{ \widehat{\chi}_n,  \widehat{h}_\kappa\right\} +  v_{T_s}v_{\parallel}\frac{\partial \widehat{h}_l}{\partial z} = \frac{q_s}{T_s}  \frac{ e^{-v^2_{\parallel}}}{\pi^{3/2}}  \frac{\partial \widehat{\chi}_l}{\partial t}\,.
\end{equation}

With the GK equation expressed in Laguerre spectral space, we can now take a  drift kinetic long wavelength limit, while retaining FLR effects. We do this in the next section and obtain the equations of interests.

\section{Drift-kinetic system of equations} \label{section:Drift-kinetic limit}

\subsection{Drift-kinetic limit}

\begin{figure}
\centering 
\includegraphics[width=0.95\textwidth]{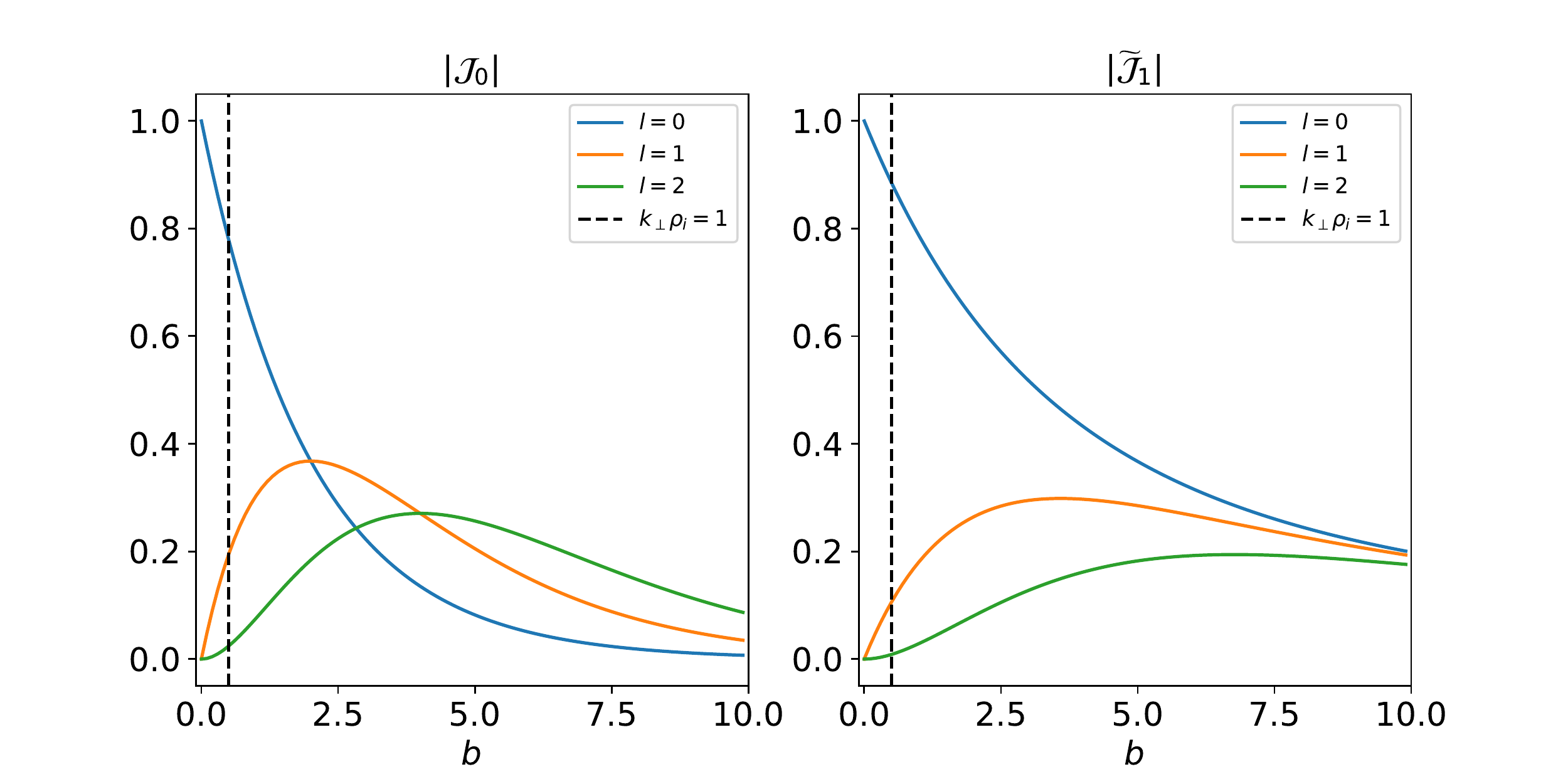}
\caption{Absolute value of moments $\mathcal{J}_{0l}$ and $\mathcal{\tilde{J}}_{1l}$. With $b\rightarrow 0$, all the higher moments approach zero. }
\label{figure:bessel}
\end{figure}

In a drift-kinetic approach, the long-wavelength limit $k_\perp \rho_s\rightarrow 0$ leads to $b_s\rightarrow 0$. Dispensing with the species label, for $b\rightarrow 0$ all $l\ne0$ terms in (\ref{equation:J round 0}-\ref{equation:J round 1}) will tend towards zeros (see Figure \ref{figure:bessel}). The only non-zero contributions, for $l=0$, are responsible for retaining FLR effects and have the explicit form,
\begin{align}
\mathcal{J}_{00} &= e^{-b/2}\,,\\
\mathcal{\tilde{J}}_{10} &= \left[\left(1-e^{-b/2}\right)\frac{2}{b}\right]\,.
\end{align}

To obtain our equations, which retain the dominant FLR contributions, we set $\mathcal{J}_{0l}=0$ and $\mathcal{\tilde{J}}_{1l}=0$ for all $l>0$ in the GK equations expressed in Laguerre form. By doing so, the field equations \eqref{phil}-\eqref{Bl} now give
\begin{align} 
        \phi(\mathbf{k},t) &= \pi\sum_s  q_s n_s\int^{+\infty}_{-\infty}dv_\parallel\mathcal{J}_{00} \widehat{h}_{s0} \bigg{/} \sum_s\frac{q_s^2 n_s}{T_s}\,,\label{eq:philTruncated}\\
    A_\parallel(\mathbf{k},t) &= \frac{\pi \beta}{2 k_\perp^2}\sum_s  q_s n_s v_{T_s}\int^{+\infty}_{-\infty}dv_\parallel v_\parallel \mathcal{J}_{00} \widehat{h}_{s0}   \,, \label{eq:AlTruncated}\\
B_\parallel(\mathbf{k},t) &= -\frac{\pi\beta}{2}\sum_s \frac{n_s T_s}{B_0}\int^{+\infty}_{-\infty}dv_\parallel \tilde{\mathcal{J}}_{10} \left[\widehat{h}_{s0} +\widehat{h}_{s1} \right] \,.\label{eq:BlTruncated}
\end{align}
Looking next at the gyrokinetic potential \eqref{chil}, we obtain $\widehat \chi_{0}$ and $\widehat \chi_{1}$, 
\begin{align}
\widehat{\chi}_{0} &= \mathcal{J}_{00}\left(\phi(\mathbf{k})-v_{T_s} v_\parallel  A_\parallel(\mathbf{k})\right) + \frac{T_s}{q_s B_0} B_\parallel(\mathbf{k})
   \tilde{\mathcal{J}}_{10}\label{eq:chiexpl0}\,,\\
\widehat{\chi}_1 &= \frac{T_s}{q_s B_0} B_\parallel(\mathbf{k}) \tilde{\mathcal{J}}_{10}\,,\label{eq:chiexpl1}
\end{align}
with all other contributions ($l>1$) being zero. 

We now see that while eqs.~(\ref{eq:philTruncated}-\ref{eq:AlTruncated}) only include the zeroth Laguerre moment of the distribution function $h$, eq.\eqref{eq:BlTruncated} also depends on the first moment. Therefore, one can solve the gyrokinetic system for only the first two moments of the distribution function. The only apparent problem is due to the nonlinear terms, which for  $l=0$ equations are given as $\frac{1}{B_0}\sum_{\kappa = 0}^\infty\sum_{n = |\kappa-0|}^{\kappa+0} \alpha_{0\kappa n}\{ \widehat{\chi}_n,  \widehat{h}_\kappa\}$, and for $l=1$ are found from $\frac{1}{B_0}\sum_{\kappa = 0}^\infty\sum_{n = |\kappa-1|}^{\kappa+1} \alpha_{1\kappa n} \{ \widehat{\chi}_n,  \widehat{h}_\kappa\}$. However, since for $\widehat{\chi}_n$ in the nonlinear terms only $n=0$ and $n=1$ terms are nonzero, and computing the required convolution coefficients $\alpha_{0\kappa n}$ and $\alpha_{1\kappa n}$ from \eqref{nonlinearl}, we obtain that the nonlinear terms are expressed solely in term of $\widehat{h}_0$ and $\widehat{h}_1$,
\begin{align}
   & \frac{\partial \widehat{h}_0}{\partial t} + \frac{1}{B_0}\left(\left\{ \widehat{\chi}_0,  \widehat{h}_0\right\}+\left\{ \widehat{\chi}_1,  \widehat{h}_1\right\}\right)
     +  v_{T_s}v_{\parallel}\frac{\partial \widehat{h}_0}{\partial z} = \frac{q_s}{T_s}  \frac{ e^{-v^2_{\parallel}}}{\pi^{3/2}}  \frac{\partial \widehat{\chi}_0}{\partial t}\label{Vlasovl0expl}\\
 &    \frac{\partial \widehat{h}_1}{\partial t} + \frac{1}{B_0}\left(\left\{ \widehat{\chi}_0,  \widehat{h}_1\right\}+\left\{ \widehat{\chi}_1,  \widehat{h}_0\right\}+ 2\left\{ \widehat{\chi}_1,  \widehat{h}_1\right\}\right) +  v_{T_s}v_{\parallel}\frac{\partial \widehat{h}_1}{\partial z} = \frac{q_s}{T_s}  \frac{ e^{-v^2_{\parallel}}}{\pi^{3/2}}  \frac{\partial \widehat{\chi}_1}{\partial t}\label{Vlasovl1expl}
\end{align}

We see that we obtain a closed set of equations. 
With $\widehat{h}_0=\widehat{h}_0({\bf R},v_\|)$ and $\widehat{h}_1=\widehat{h}_1({\bf R},v_\|)$ obtained explicitly as the largest $\mu$ space scale-structure, our four-dimensional set of equations retain FLR effects. The equations themselves serve to capture electromagnetic plasma turbulence in the drift kinetic limit, while allowing to transition into the sub-gyroradius range of scales captured by GK theory. Our equations account for the linear phase mixing that occurs in $v_\|$ space due to the particle-field resonance. With an interest in the interplay between spatial turbulence and $v_\|$ space mixing, we will perform in the next section a Hermite representation of our equations in the $v_\|$ direction and a Fourier representation for spatial scales. We first comment on the conservation of free energy and the electrostatic approximation.

\subsection{Free energy conservation}

Following the works of \cite{Howes:2006p1280, Schekochihin_2009}, in \cite{Teaca:SGS2021} we took the ${\bf R}$-density for the free energy contribution of a species $s$ to be $W_s({\bf R},t)=  \pi \int dv_\| d\mu B_0 \big{[}h_s- \frac{q_s F_s}{T_s} \chi \,\big{]}\frac{T_s}{F_s}h_s$. Dropping the species label, using for our drift kinetic system $g_0 = h_0- \frac{q_s}{T_s} { e^{-v^2_{\parallel}}}{\pi^{-3/2}}  \chi_0$ and $g_1 = h_1- \frac{q_s}{T_s} { e^{-v^2_{\parallel}}}{\pi^{-3/2}}  \chi_1$ , we obtain for in non-dimensional form $W({\bf R},t)=  \pi^{5/2} T \int dv_\|  e^{{v_\parallel^2}} \big{[}\widehat h_0 \widehat g_0+\widehat h_1 \widehat g_1\,\big{]}$.

Rewriting eqs.~(\ref{Vlasovl0expl}-\ref{Vlasovl1expl}) as
\begin{align}
   & \frac{\partial \widehat{g}_0}{\partial t} + \frac{1}{B_0}\left(\left\{ \widehat{\chi}_0,  \widehat{h}_0\right\}+\left\{ \widehat{\chi}_1,  \widehat{h}_1\right\}\right)
     +  v_{T_s}v_{\parallel}\frac{\partial \widehat{h}_0}{\partial z} = 0\label{g0}\,,\\
 &    \frac{\partial \widehat{g}_1}{\partial t} + \frac{1}{B_0}\left(\left\{ \widehat{\chi}_0,  \widehat{h}_1\right\}+\left\{ \widehat{\chi}_1,  \widehat{h}_0\right\}+ 2\left\{ \widehat{\chi}_1,  \widehat{h}_1\right\}\right) +  v_{T_s}v_{\parallel}\frac{\partial \widehat{h}_1}{\partial z} = 0\label{g1}\,,
\end{align}
multiplying eq. \eqref{g0} by $\widehat{h}_0$ and eq. \eqref{g1}  by $\widehat{h}_1$ and summing them we get
\begin{align}
    \frac{\partial \widehat{g}_0}{\partial t}\widehat{h}_0 + \frac{\partial \widehat{g}_1}{\partial t}  \widehat{h}_1 &+  \frac{1}{2}v_{T_s}v_{\parallel}\bigg{[} \frac{\partial \widehat{h}_0^2}{\partial z} +  \frac{\partial \widehat{h}_1^2}{\partial z} \bigg{]}+ \nonumber\\
   &\frac{1}{B_0}\bigg{[}\frac{1}{2}\left\{ \widehat{\chi}_0,  \widehat{h}^2_0\right\}+\frac{1}{2}\left\{ \widehat{\chi}_0,  \widehat{h}_1^2\right\}+ \left\{ \widehat{\chi}_1,  \widehat{h}_1^2\right\}+
   \left\{ \widehat{\chi}_1,  \widehat{h}_0\widehat{h}_1\right\}\bigg{]}  = 0\,,
\end{align}
where the Poisson bracket and the $z$ derivatives term integrate spatially to zero. We find that $\int d{\bf R}\partial W/\partial t \equiv \int d{\bf R}[ h_0\partial g_0/\partial t + h_1\partial g_1/\partial t ]=0$, which shows that the free energy $W$ is conserved globally in the absence of collisions or external energy sources in the system.

\subsection{Electrostatic case}

The electrostatic limit ($\beta \rightarrow 0$) represents a special case of interest for our drift-kinetic equations. It can be used to compare with other drift-kinetic models derived solely for electrostatic fluctuations.

With $A_\parallel$ and $B_\parallel$ fields vanishing, there is no need to keep first Laguerre moment ($h_1^m$) and its associate equations. Therefore, our drift kinetic equations reduce to the one describing the evolution of the zeroth Laguerre moment,
\begin{align}
    \frac{\partial \widehat{h}_0}{\partial t} + \frac{1}{B_0} \left\{ \widehat{\chi}_0,  \widehat{h}_0\right\} +  v_{T_s}v_{\parallel}\frac{\partial \widehat{h}_0}{\partial z} &= \frac{q}{T} \frac{ e^{-v^2_{\parallel}}}{\pi^{3/2}} \frac{\partial \widehat{\chi}_0}{\partial t}
\end{align}
with  
\begin{align}
\widehat{\chi}_{0} &= \mathcal{J}_{00}\phi(\mathbf{k})\,,\\
 \phi(\mathbf{k},t) &= \pi\sum_s  q_s n_s\int^{+\infty}_{-\infty}dv_\parallel\mathcal{J}_{00} \widehat{h}_{s0} \bigg{/} \sum_s\frac{q_s^2 n_s}{T_s} \label{philTruncated}\,.
\end{align}

Going further and selecting an adiabatic species, can lead to a further simplified model as presented in \cite{hatch2014} in a Fourier-Hermite spectral form. In the next section we present the Fourier-Hermite spectral form for our system.

\section{Drift-kinetic system of equations in spectral form}\label{section:spectraleq}

To finish writing our drift kinetic equation is spectral form, we will utilise a Hermite decomposition for $v_\|$ and a Fourier decomposition in $\bf R$. We first perform the Hermite representation. 

\subsection{Hermite basis projection} \label{section:Hermite transform}

For convenience, we chose the Hermite function basis used by \cite{mandell2018}.The Hermite functions in question are given as 
\begin{align}
\xi^m\left(v_\parallel\right) &= \frac{(-1)^m }{(2^m m! \pi^3)^{1/2}} e^{-v^2_\parallel} He(v_\parallel) \,,\\
\xi_m\left(v_\parallel\right) &= \frac{(-1)^m }{(2^m m!)^{1/2}} He(v_\parallel)\,,
\end{align}
where $He(v_\parallel) = e^{v_\parallel^2}\frac{d^m}{d v_\parallel^m} e^{-v_\parallel^2}$ are Hermite polynomials. The Hermite functions obey the orthogonality condition
\begin{equation}\label{eq:orthHermite}
   \pi \int_{-\infty}^{+\infty} dv_\parallel \xi^m(v_\parallel) \xi_n(v_\parallel) = \delta_{nm} \;,
\end{equation}
and, among others, the recurrence relation 
\begin{equation}\label{eq:recHermite}
    v_\parallel\xi^m(v_\parallel) = \sqrt{\frac{m}{2}}\xi^{m-1}(v_\parallel) + \sqrt{\frac{m+1}{2}}\xi^{m+1}(v_\parallel) \,.
\end{equation}

The decomposition on the Hermite basis is done in the same manner as it was done for the Laguerre basis. Function in Hermite basis can be represented as 
\begin{equation}\label{eq:hermiteRepresentation}
    f(v_\parallel) = \sum_{m=0}^{\infty} \xi^m(v_\parallel) f^m \,,
\end{equation}
with spectral amplitude 
\begin{equation}\label{eq:hermiteAmplitude}
    f^m = \pi\int_{-\infty}^{+\infty} dv_\parallel \xi_m(v_\parallel) f(v_\parallel)\,.
\end{equation}

In Hermite space, by using $\psi_0(v_\|)=1$ and the recurrence formula \eqref{eq:recHermite}, the field equations are found to be
\begin{align}
\phi(\mathbf{k},t) &= \pi\sum_s  q_s n_s\int^{+\infty}_{-\infty}dv_\parallel  \sum_{m=0}^{\infty}\mathcal{J}_{00}\xi^m(v_\|)\widehat{h}^m_{s0} \bigg{/} \sum_s\frac{q_s^2 n_s}{T_s} \nonumber\\
&= \sum_s  q_s n_s \mathcal{J}_{00}\widehat{h}^0_{s0} \bigg{/} \sum_s\frac{q_s^2 n_s}{T_s}\,,\label{eq:phiHermite}\\
A_\parallel(\mathbf{k},t) &= \frac{\pi \beta}{2 k_\perp^2}\sum_s  q_s n_s v_{T_s}\int^{+\infty}_{-\infty}dv_\parallel v_\parallel \mathcal{J}_{00} \sum_{m=0}^{\infty}\xi^m(v_\parallel) \widehat{h}^m_{s0} \nonumber\\
            &= \frac{\pi \beta}{2 k_\perp^2}\sum_s  q_s n_s v_{T_s}\int^{+\infty}_{-\infty}dv_\parallel \mathcal{J}_{00} \sum_{m=0}^{\infty}\left(\sqrt{\frac{m}{2}}\xi^{m-1}(v_\parallel)+\sqrt{\frac{m+1}{2}}\xi^{m+1}(v_\parallel)\right)\widehat{h}^m_{s0} \nonumber\\
            &= \frac{ \beta}{2 k_\perp^2}\sum_s  q_s n_s v_{T_s}\mathcal{J}_{00} \sqrt{\frac{1}{2}} \widehat{h}^1_{s0}\,,\label{eq:AparHermite}\\
 B_\parallel(\mathbf{k},t) &= -\frac{\beta}{2}\sum_s \frac{n_s T_s}{B_0} 
  \tilde{\mathcal{J}}_{10} \left(\widehat{h}^0_{s0} + \widehat{h}^0_{s1} \right) \,,\label{eq:BparHermite}
\end{align}
where for $B_\parallel$ the procedure is the same as for \eqref{eq:phiHermite}. To help with the representation of the GK equation in Hermite space, as the $A_\|$ contribution contains a $v_\|$ factor, and the $B_\|$ contribution is the same in \eqref{eq:chiexpl0} and \eqref{eq:chiexpl1}, we represent $\widehat \chi_0$ and $\widehat \chi_1$ as
\begin{align}
    &\widehat{\chi}_0 = \widehat{\chi}^{\phi} + \widehat{\chi}^{B} +  v_\parallel  \widehat{\chi}^{A }\label{eq:chiCompositionHermite}\,,\\
    &\widehat{\chi}_1 = \widehat{\chi}^{B} \,,
\end{align}
with
\begin{align}
    \widehat{\chi}^{\phi}(\mathbf{k},t) &= \mathcal{J}_{00}\phi(\mathbf{k},t) \,,\\
    \widehat{\chi}^{B}(\mathbf{k},t) &=  \frac{T}{q B_0} \tilde{\mathcal{J}}_{10}B_\parallel(\mathbf{k},t) \,,\\
    \widehat{\chi}^{A}(\mathbf{k},t) &= - v_{T_s} \mathcal{J}_{00} A_\parallel(\mathbf{k},t)\,.
\end{align}

The projection of the GK equation on the Hermite basis is done in straightforward manner. One has to use \eqref{eq:hermiteRepresentation} for $h_0$ and $h_1$, the recurrence formula \eqref{eq:recHermite} for terms that contain $v_\|$, and the integrate over $\pi\int_{-\infty}^{+\infty} d v_\parallel \xi_m$. As an example we look at the nonlinear term $\frac{1}{B_0}\{ \chi_0, h_0\}$, which becomes
\begin{align}
        \frac{\pi}{B_0}\int_{-\infty}^{+\infty} d v_\parallel \xi_m&\left\{ \widehat{\chi}^{\phi}+\widehat{\chi}^{B}  +  v_\parallel  \widehat{\chi}^{A} ,  \sum_n \xi^n \widehat{h}^n_0\right\} = \frac{1}{B_0}\left\{ \widehat{\chi}^{\phi}, \widehat{h}^m_0\right\}
        +\frac{1}{B_0}\left\{ \widehat{\chi}^{B}, \widehat{h}^m_0\right\} \nonumber\\
        &+ \frac{1}{B_0} \sqrt{\frac{m+1}{2}}\left\{\widehat{\chi}^{A} ,\widehat{h}_0^{m+1}\right\}
        +\frac{1}{B_0} \sqrt{\frac{m}{2}}\left\{\widehat{\chi}^{A} ,\widehat{h}^{m-1}_0\right\}\,.
\end{align}

The GK equation expressed in Hermite space have the form
\begin{align}
\frac{\partial{\widehat{h}}^m_0}{\partial t} &+ \frac{1}{B_0}\Bigg{[}\left\{ \widehat{\chi}^{\phi}+\widehat{\chi}^{B}, \widehat{h}^m_0\right\} + \sqrt{\frac{m+1}{2}}\left\{\widehat{\chi}^{A} ,\widehat{h}_0^{m+1}\right\}+\sqrt{\frac{m}{2}}\left\{\widehat{\chi}^{A} ,\widehat{h}^{m-1}_0\right\}+\left\{ \widehat{\chi}^{B}, \widehat{h}^m_1\right\} \Bigg{]} \nonumber\\
&+v_{T_s} \frac{\partial }{\partial z}\Bigg{[} \sqrt{\frac{m+1}{2}}\widehat{h}^{m+1}_0+\sqrt{\frac{m}{2}}\widehat{h}^{m-1}_0 \Bigg{]}= \frac{q}{T} \frac{\partial }{\partial t} \Bigg{[}( \widehat{\chi}^{\phi} +\widehat{\chi}^{B}) \delta_{m0}+\sqrt{\frac{1}{2}} \widehat{\chi}^{A}\delta_{m,1} \Bigg{]}\label{eq:Vlasov0Hermite}\\
\frac{\partial h^m_1}{\partial t} &+ \frac{1}{B_0}\Bigg{[}\left\{ \widehat{\chi}^{\phi}+\widehat{\chi}^{B}, \widehat{h}^m_1\right\} + \sqrt{\frac{m+1}{2}}\left\{\widehat{\chi}^{A} ,\widehat{h}_1^{m+1}\right\} +\sqrt{\frac{m}{2}}\left\{\widehat{\chi}^{A} ,\widehat{h}^{m-1}_1\right\}+\left\{ \widehat{\chi}^{B}, \widehat{h}^m_0\right\}\nonumber\\
&+2\left\{ \widehat{\chi}^{B}, \widehat{h}^m_1\right\} \Bigg{]}+v_{T_s} \frac{\partial }{\partial z}\Bigg{[} \sqrt{\frac{m+1}{2}}\widehat{h}^{m+1}_1+\sqrt{\frac{m}{2}}\widehat{h}^{m-1}_1 \Bigg{]}=\frac{q}{T} \frac{\partial \widehat{\chi}^{B}}{\partial t}\delta_{m,0}\label{eq:Vlasov1Hermite}
\end{align}

Expressing these equation in Fourier space is trivial and we will simply present them in the next section.

\subsection{Drift-kinetic system in spectral space}

To keep notations as simple as possible, we write the Fourier space representation for the Poisson bracket between two functions $f$ and $g$ as 
\begin{align}
{\{f,g\}}({\bf k}) = \frac{1}{2}\sum_{\mathbf{p}+\mathbf{q}=\mathbf{k}}\left(q_x p_y - q_y p_x\right)&\Big[\widehat f({\bf q}) \widehat g({\bf p})- \widehat f({\bf p}) \widehat g({\bf q}) \Big]\,.  \label{eq:nonlinearTermFourier0}
\end{align}

Dispensing with the $\widehat{...}$ designation for (Fourier-Hermite-Laguerre space) modes to further simplify notations, the drift-kinetic equations in spectral form have the form 
\begin{align}
\frac{\partial{g}^m_0}{\partial t} &+ \frac{1}{B_0}\Bigg{[}\left\{ \chi^{\phi}+\chi^{B}, h^m_0\right\} + \sqrt{\frac{m+1}{2}}\left\{\chi^{A} ,h_0^{m+1}\right\}+\sqrt{\frac{m}{2}}\left\{\chi^{A} ,h^{m-1}_0\right\}\nonumber\\
&+\left\{ \chi^{B}, h^m_1\right\} \Bigg{]}  +ik_z v_{T_s} \Bigg{[} \sqrt{\frac{m+1}{2}}h^{m+1}_0+\sqrt{\frac{m}{2}}h^{m-1}_0 \Bigg{]}= 0 \\
\frac{\partial g^m_1}{\partial t} &+ \frac{1}{B_0}\Bigg{[}\left\{ \chi^{\phi}+\chi^{B}, h^m_1\right\} + \sqrt{\frac{m+1}{2}}\left\{\chi^{A} ,h_1^{m+1}\right\} +\sqrt{\frac{m}{2}}\left\{\chi^{A} ,h^{m-1}_1\right\}\nonumber\\
&+\left\{ \chi^{B}, h^m_0\right\}+2\left\{ \chi^{B}, h^m_1\right\} \Bigg{]}+ik_z v_{T_s} \Bigg{[} \sqrt{\frac{m+1}{2}}h^{m+1}_1+\sqrt{\frac{m}{2}}h^{m-1}_1 \Bigg{]}=0
\end{align}
where it is convenient to use the modified gyrokinetic distribution functions,
\begin{align}
    &g^m_0 = h^m_0 - \frac{q}{T}(\chi^{\phi} +\chi^{B})\delta_{m0} -  \frac{q}{T}\sqrt{\frac{1}{2}}\chi^{A} \delta_{m1}\\
    &g^m_1 = h^m_1 - \frac{q}{T} \chi^{B}\delta_{m0}
\end{align}
with the GK potential contributions given as
\begin{align}
    \chi^{\phi}(\mathbf{k}) &= \mathcal{J}_{00}\phi(\mathbf{k}) \,,\\
    \chi^{B}(\mathbf{k}) &=  \frac{T}{q B_0} \tilde{\mathcal{J}}_{10}B_\parallel(\mathbf{k}) \,,\\
    \chi^{A}(\mathbf{k}) &= - v_{T_s} \mathcal{J}_{00} A_\parallel(\mathbf{k})\,.
\end{align}
for fields 
\begin{align}
\phi(\mathbf{k},t) &= \sum_s  q_s n_s \mathcal{J}_{00}h^0_{s0} \bigg{/} \sum_s\frac{q_s^2 n_s}{T_s}\,,\label{eq:phiHermite2}\\
A_\parallel(\mathbf{k},t) &= \frac{ \beta}{2 k_\perp^2}\sum_s  q_s n_s v_{T_s} \mathcal{J}_{00}\sqrt{\frac{1}{2}} h^1_{s0}\,,\label{eq:AparHermite2}\\
 B_\parallel(\mathbf{k},t) &= -\frac{\beta}{2}\sum_s \frac{n_s T_s}{B_0} 
  \tilde{\mathcal{J}}_{10} \left[h^0_{s0} +h^0_{s1} \right] \,,\label{eq:BparHermite2}
\end{align}
and Bessel function contributions that account for FLR effects,
\begin{align}
   \mathcal{J}_{00} &= e^{-b/2}\label{equation:J round 00}\,,\\
   \mathcal{\tilde{J}}_{10} &= \left[\left(1-e^{-b/2}\right)\frac{2}{b}\right]\,.\label{equation:J round 11}
\end{align}

These are the equations that will be solved numerically in a pseudo-spectral form or in a triad reduced model for the nonlinear terms. We have not include a collision operator in our presentation. One can chose to use the one presented in \cite{mandell2018}, or simply use a Laplacian or hyper-Laplacian expressed locally in $m$ and $k_\perp$ space as means to provide a sink of free energy at small scales. However, in the collisionless regimes, the dissipation of free energy from a truncated range of scales is due to an anomalous dissipation generated by the flux of free energy to the unresolved scales \cite{PhysRevX.8.041020}. With this discussion being lengthly and requiring numerical results to support it, we choose not to delve in the presentation of any dissipative mechanism.

\section{Conclusions}

The goal of this article was to derive a set of four-dimensional drift kinetic equations that conserve free energy, and are capable of describing electromagnetic turbulence in a magnetised plasma at the transition between fluid and GK regimes ($k_\perp \rho_i\approx 1$ range of scales). Starting from the GK equations, we have employed a Laguerre decomposition of the distribution functions in the perpendicular velocity direction and have retained only the first two Laguerre moments that source the electromagnetic fluctuations. FLR effects are kept via the dominant contributions of the Bessel functions, expanded in a Laguerre polynomial basis, allowing for the transition into the GK range of scales to be possible.

The Laguerre-Hermite representation has been used in \cite{mandell2018} to present a spectral approach for electrostatic GK in toroidal magnetic geometry. Here, not only that we expanded this to electromagnetic fluctuations, we have have effectively integrated out the $\mu$ dependence to reduce the dimensions of the kinetic system at $k_\perp\rho_i <1$. In addition to the three-dimensions in positions space, we retained the parallel velocity dependence, which we captured via a Hermite representation. Employing this system, but without any other physics based assumptions for the plasma species, will allow us to investigate how fluid effects transition into the kinetic range in the astrophysical context.

Currently, there are a large number of models describing the astrophysical plasma in the large span of scales natural for such system \cite{Schekochihin_2009, zocco2011, adkins_2018}. In general, they all can be combined into three big groups: (i) kinetic models that take into account all the kinetic effects,  (ii) MHD models that treat plasma as a fluid of charged particles, (iii) and kinetic-reduced MHD (KRMHD) models \cite{kunz2018, Meyrand2019} that assume isothermal (fluid) electrons, which are applicable for scales smaller or approximately to that of the ion gyroradius. The latter models provide the computational effectiveness, and are able to capture some important kinetic effects. However, since most of those models utilise additional physical assumptions to eliminate one of the species, some of the kinetic effects can fall out of the scope of these models, or they can be captured in a very slim way. Therefore, our choice was to use the route highlighted by \cite{brizard1992} and derive a more comprehensive drift-kinetic model, in order to track down kinetic effects arising at the border of transition from fluid to (gyro)kinetic turbulence. Our approach is better suited to the analysis of the interplay between spatial and velocity space mixing for electromagnetic plasma turbulence, as no a-priori assumption on the fluid nature of a species is made. This will be pursued numerically via pseudo-spectral solvers and low dimensional models.
 
\section*{Acknowledgements} BT would like to thank Gabriel Plunk and David Hatch for discussions on GK dynamics and reduced models.


\begin{thebibliography}{42}
\expandafter\ifx\csname natexlab\endcsname\relax\def\natexlab#1{#1}\fi
\def\au#1{#1} \def\ed#1{#1} \def\yr#1{#1}\def\at#1{#1}\def\jt#1{\textit{#1}}
  \def\bt#1{#1}\def\bvol#1{\textbf{#1}} \def\vol#1{#1} \def\pg#1{#1}
  \def\publ#1{#1}\def\arxiv#1{#1}\def\org#1{#1}\def\st#1{\textit{#1}}

\bibitem[Adkins \& Schekochihin(2018)]{adkins_2018}
{\sc \au{Adkins, T.} \& \au{Schekochihin, A.~A.}} \yr{2018}  \at{A solvable
  model of vlasov-kinetic plasma turbulence in fourier–hermite phase space}.
  \jt{Journal of Plasma Physics}  \bvol{84}~(1),  \pg{905840107}.

\bibitem[Alexandrova {\em et~al.\/}(2008)Alexandrova, Carbone, Veltri \&
  Sorriso-Valvo]{Alexandrova:2008p1397}
{\sc \au{Alexandrova, O.}, \au{Carbone, V.}, \au{Veltri, P.} \&
  \au{Sorriso-Valvo, L.}} \yr{2008}  \at{Small-scale energy cascade of the
  solar wind turbulence}.  \jt{The Astrophysical Journal}  \bvol{674},
  \pg{1153}.

\bibitem[Alexandrova {\em et~al.\/}(2009)Alexandrova, Saur, Lacombe, Mangeney,
  Mitchell, Schwartz \& Robert]{Alexandrova:2009p1396}
{\sc \au{Alexandrova, O.}, \au{Saur, J.}, \au{Lacombe, C.}, \au{Mangeney, A.},
  \au{Mitchell, J.}, \au{Schwartz, S.~J.} \& \au{Robert, P.}} \yr{2009}
  \at{Universality of solar-wind turbulent spectrum from mhd to electron
  scales}.  \jt{Phys. Rev. Lett.}  \bvol{103},  \pg{165003}.

\bibitem[Bhattacharjee {\em et~al.\/}(1998)Bhattacharjee, Ng \&
  Spangler]{Bhattacharjee:1998p313}
{\sc \au{Bhattacharjee, A.}, \au{Ng, C.~S.} \& \au{Spangler, S.~R.}} \yr{1998}
  \at{Weakly compressible magnetohydrodynamic turbulence in the solar wind and
  the interstellar medium}.  \jt{The Astrophysical Journal}  \bvol{494},
  \pg{409}.

\bibitem[Brizard(1992)]{brizard1992}
{\sc \au{Brizard, A.}} \yr{1992}  \at{Nonlinear gyrofluid description of
  turbulent magnetized plasmas}.  \jt{Physics of Fluids B: Plasma Physics}
  \bvol{4}~(5),  \pg{1213--1228}.

\bibitem[Cerri {\em et~al.\/}(2018)Cerri, Kunz \& Califano]{Cerri:2018p2139}
{\sc \au{Cerri, S.~S.}, \au{Kunz, M.~W.} \& \au{Califano, F.}} \yr{2018}
  \at{Dual phase-space cascades in 3d hybrid-vlasov-maxwell turbulence}.
  \jt{The Astrophysical Journal Letters}  \bvol{856},  \pg{L13}.

\bibitem[Chandran {\em et~al.\/}(2009)Chandran, Quataert, Howes, Hollweg \&
  Dorland]{Chandran:2009p760}
{\sc \au{Chandran, B.}, \au{Quataert, E.}, \au{Howes, G.~G.}, \au{Hollweg,
  J.~V.} \& \au{Dorland, W.}} \yr{2009}  \at{The turbulent heating rate in
  strong magnetohydrodynamic turbulence with nonzero cross helicity}.  \jt{The
  Astrophysical Journal}  \bvol{701},  \pg{652}.

\bibitem[Dorland \& Hammett(1993)]{Dorland93}
{\sc \au{Dorland, W.} \& \au{Hammett, G.~W.}} \yr{1993}  \at{Gyrofluid
  turbulence models with kinetic effects}.  \jt{Physics of Fluids B: Plasma
  Physics}  \bvol{5}~(3),  \pg{812--835}.

\bibitem[Duan {\em et~al.\/}(2020)Duan, Bowen, Chen, Mallet, He, Bale, Vech,
  Kasper, Pulupa, Bonnell, Case, de~Wit, Goetz, Harvey, Korreck, Larson, Livi,
  MacDowall, Malaspina, Stevens \& Whittlesey]{Duan_2020}
{\sc \au{Duan, D.}, \au{Bowen, T.~A.}, \au{Chen, C. H.~K.}, \au{Mallet, A.},
  \au{He, J.}, \au{Bale, S.~D.}, \au{Vech, D.}, \au{Kasper, J.~C.}, \au{Pulupa,
  M.}, \au{Bonnell, J.~W.}, \au{Case, A.~W.}, \au{de~Wit, T.~D.}, \au{Goetz,
  K.}, \au{Harvey, P.~R.}, \au{Korreck, K.~E.}, \au{Larson, D.}, \au{Livi, R.},
  \au{MacDowall, R.~J.}, \au{Malaspina, D.~M.}, \au{Stevens, M.} \&
  \au{Whittlesey, P.}} \yr{2020}  \at{The radial dependence of proton-scale
  magnetic spectral break in slow solar wind during {PSP} encounter 2}.
  \jt{The Astrophysical Journal Supplement Series}  \bvol{246}~(2),  \pg{55}.

\bibitem[Duan {\em et~al.\/}(2021)Duan, He, Bowen, Woodham, Wang, Chen, Mallet
  \& Bale]{Duan_2021}
{\sc \au{Duan, D.}, \au{He, J.}, \au{Bowen, T.~A.}, \au{Woodham, L.~D.},
  \au{Wang, T.}, \au{Chen, C. H.~K.}, \au{Mallet, A.} \& \au{Bale, S.~D.}}
  \yr{2021}  \at{Anisotropy of solar wind turbulence in the inner heliosphere
  at kinetic scales: {PSP} observations}.  \jt{The Astrophysical Journal
  Letters}  \bvol{915}~(1),  \pg{L8}.

\bibitem[Eyink(2018)]{PhysRevX.8.041020}
{\sc \au{Eyink, G.~L.}} \yr{2018}  \at{Cascades and dissipative anomalies in
  nearly collisionless plasma turbulence}.  \jt{Phys. Rev. X}  \bvol{8},
  \pg{041020}.

\bibitem[Goldreich \& Sridhar(1995)]{Goldreich:1995p724}
{\sc \au{Goldreich, P.} \& \au{Sridhar, S.}} \yr{1995}  \at{Toward a theory of
  interstellar turbulence. 2: Strong alfvenic turbulence}.  \jt{The
  Astrophysical Journal}  \bvol{438},  \pg{763}.

\bibitem[Gro{\v s}elj {\em et~al.\/}(2019)Gro{\v s}elj, Chen, Mallet, Samtaney,
  Schneider \& Jenko]{2019PhRvX...9c1037G}
{\sc \au{Gro{\v s}elj, D.}, \au{Chen, C. H.~K.}, \au{Mallet, A.}, \au{Samtaney,
  R.}, \au{Schneider, K.} \& \au{Jenko, F.}} \yr{2019}  \at{{Kinetic Turbulence
  in Astrophysical Plasmas: Waves and/or Structures?}}  \jt{Physical Review X}
  \bvol{9}~(3),  \pg{031037}.

\bibitem[Hatch {\em et~al.\/}(2014)Hatch, Jenko, Bratanov \&
  Navarro]{hatch2014}
{\sc \au{Hatch, D.~R.}, \au{Jenko, F.}, \au{Bratanov, V.} \& \au{Navarro,
  A.~B.}} \yr{2014}  \at{Phase space scales of free energy dissipation in
  gradient-driven gyrokinetic turbulence}.  \jt{Journal of Plasma Physics}
  \bvol{80}~(4),  \pg{531–551}.

\bibitem[Howes {\em et~al.\/}(2006)Howes, Cowley, Dorland, Hammett, Quataert \&
  Schekochihin]{Howes:2006p1280}
{\sc \au{Howes, G.~G.}, \au{Cowley, S.~C.}, \au{Dorland, W.}, \au{Hammett,
  G.~W.}, \au{Quataert, E.} \& \au{Schekochihin, A.~A.}} \yr{2006}
  \at{Astrophysical gyrokinetics: Basic equations and linear theory}.  \jt{The
  Astrophysical Journal}  \bvol{651},  \pg{590}.

\bibitem[Howes {\em et~al.\/}(2008)Howes, Dorland, Cowley, Hammett, Quataert,
  Schekochihin \& Tatsuno]{Howes:2008p1132}
{\sc \au{Howes, G.~G.}, \au{Dorland, W.}, \au{Cowley, S.~C.}, \au{Hammett,
  G.~W.}, \au{Quataert, E.}, \au{Schekochihin, A.~A.} \& \au{Tatsuno, T.}}
  \yr{2008}  \at{Kinetic simulations of magnetized turbulence in astrophysical
  plasmas}.  \jt{Phys. Rev. Lett.}  \bvol{100},  \pg{65004}.

\bibitem[Howes {\em et~al.\/}(2011)Howes, Tenbarge, Dorland, Quataert,
  Schekochihin, Numata \& Tatsuno]{Howes:2011p1370}
{\sc \au{Howes, G.~G.}, \au{Tenbarge, J.~M.}, \au{Dorland, W.}, \au{Quataert,
  E.}, \au{Schekochihin, A.~A.}, \au{Numata, R.} \& \au{Tatsuno, T.}} \yr{2011}
   \at{Gyrokinetic simulations of solar wind turbulence from ion to electron
  scales}.  \jt{Phys. Rev. Lett.}  \bvol{107},  \pg{35004}.

\bibitem[Kawazura {\em et~al.\/}(2019)Kawazura, Barnes \&
  Schekochihin]{Kawazura771}
{\sc \au{Kawazura, Y.}, \au{Barnes, M.} \& \au{Schekochihin, A.~A.}} \yr{2019}
  \at{Thermal disequilibration of ions and electrons by collisionless plasma
  turbulence}.  \jt{Proceedings of the National Academy of Sciences}
  \bvol{116}~(3),  \pg{771--776}.

\bibitem[Kunz {\em et~al.\/}(2018)Kunz, Abel, Klein \& Schekochihin]{kunz2018}
{\sc \au{Kunz, M.}, \au{Abel, I.}, \au{Klein, K.} \& \au{Schekochihin, A.}}
  \yr{2018}  \at{Astrophysical gyrokinetics: turbulence in pressure-anisotropic
  plasmas at ion scales and beyond}.  \jt{Journal of Plasma Physics}
  \bvol{84}~(2),  \pg{715840201}.

\bibitem[Mandell {\em et~al.\/}(2018)Mandell, Dorland \&
  Landreman]{mandell2018}
{\sc \au{Mandell, N.}, \au{Dorland, W.} \& \au{Landreman, M.}} \yr{2018}
  \at{Laguerre–hermite pseudo-spectral velocity formulation of gyrokinetics}.
   \jt{Journal of Plasma Physics}  \bvol{84}~(1),  \pg{905840108}.

\bibitem[Meyrand {\em et~al.\/}(2019)Meyrand, Kanekar, Dorland \&
  Schekochihin]{Meyrand2019}
{\sc \au{Meyrand, R.}, \au{Kanekar, A.}, \au{Dorland, W.} \& \au{Schekochihin,
  A.~A.}} \yr{2019}  \at{Fluidization of collisionless plasma turbulence}.
  \jt{Proceedings of the National Academy of Sciences}  \bvol{116}~(4),
  \pg{1185--1194}.

\bibitem[Meyrand {\em et~al.\/}(2021)Meyrand, Squire, Schekochihin \&
  Dorland]{meyrand2021}
{\sc \au{Meyrand, R.}, \au{Squire, J.}, \au{Schekochihin, A.} \& \au{Dorland,
  W.}} \yr{2021}  \at{On the violation of the zeroth law of turbulence in space
  plasmas}.  \jt{Journal of Plasma Physics}  \bvol{87}~(3),  \pg{535870301}.

\bibitem[Navarro {\em et~al.\/}(2016)Navarro, Teaca, Told, Groselj, Crandall \&
  Jenko]{Navarro:2016p1965}
{\sc \au{Navarro, A.~B.}, \au{Teaca, B.}, \au{Told, D.}, \au{Groselj, D.},
  \au{Crandall, P.} \& \au{Jenko, F.}} \yr{2016}  \at{Structure of plasma
  heating in gyrokinetic alfv{\'e}nic turbulence}.  \jt{Phys. Rev. Lett.}
  \bvol{117}~(24),  \pg{245101}.

\bibitem[Schekochihin {\em et~al.\/}(2008)Schekochihin, Cowley, Dorland,
  Hammett, Howes, Plunk, Quataert \& Tatsuno]{Schekochihin:2008p1034}
{\sc \au{Schekochihin, A.~A.}, \au{Cowley, S.~C.}, \au{Dorland, W.},
  \au{Hammett, G.~W.}, \au{Howes, G.~G.}, \au{Plunk, G.~G.}, \au{Quataert, E.}
  \& \au{Tatsuno, T.}} \yr{2008}  \at{Gyrokinetic turbulence: a nonlinear route
  to dissipation through phase space}.  \jt{Plasma Phys. Control. Fusion}
  \bvol{50},  \pg{4024}.

\bibitem[Schekochihin {\em et~al.\/}(2009)Schekochihin, Cowley, Dorland,
  Hammett, Howes, Quataert \& Tatsuno]{Schekochihin_2009}
{\sc \au{Schekochihin, A.~A.}, \au{Cowley, S.~C.}, \au{Dorland, W.},
  \au{Hammett, G.~W.}, \au{Howes, G.~G.}, \au{Quataert, E.} \& \au{Tatsuno,
  T.}} \yr{2009}  \at{{Astrophysical} {Gyrokinetics}: {Kinetic} {and} {Fluid}
  {Turbulent} {Cascades} {in} {Magnetized} {Weakly} {Collisional} {Plasmas}}.
  \jt{The Astrophysical Journal Supplement Series}  \bvol{182}~(1),
  \pg{310--377}.

\bibitem[Schekochihin {\em et~al.\/}(2016)Schekochihin, Parker, Highcock,
  Dellar, Dorland \& Hammett]{schekochihin_2016}
{\sc \au{Schekochihin, A.~A.}, \au{Parker, J.~T.}, \au{Highcock, E.~G.},
  \au{Dellar, P.~J.}, \au{Dorland, W.} \& \au{Hammett, G.~W.}} \yr{2016}
  \at{Phase mixing versus nonlinear advection in drift-kinetic plasma
  turbulence}.  \jt{Journal of Plasma Physics}  \bvol{82}~(2),  \pg{905820212}.

\bibitem[Servidio {\em et~al.\/}(2012)Servidio, Valentini, Califano \&
  Veltri]{Servidio:2012p1845}
{\sc \au{Servidio, S.}, \au{Valentini, F.}, \au{Califano, F.} \& \au{Veltri,
  P.}} \yr{2012}  \at{Local kinetic effects in two-dimensional plasma
  turbulence}.  \jt{Phys. Rev. Lett.}  \bvol{108},  \pg{045001}.

\bibitem[Tatsuno {\em et~al.\/}(2009)Tatsuno, Dorland, Schekochihin, Plunk,
  Barnes, Cowley \& Howes]{Tatsuno:2009p1096}
{\sc \au{Tatsuno, T.}, \au{Dorland, W.}, \au{Schekochihin, A.~A.}, \au{Plunk,
  G.~G.}, \au{Barnes, M.}, \au{Cowley, S.~C.} \& \au{Howes, G.~G.}} \yr{2009}
  \at{Nonlinear phase mixing and phase-space cascade of entropy in gyrokinetic
  plasma turbulence}.  \jt{Phys. Rev. Lett.}  \bvol{103},  \pg{15003}.

\bibitem[Teaca {\em et~al.\/}(2021)Teaca, Gorbunov, Told, Bañón~Navarro \&
  Jenko]{Teaca:SGS2021}
{\sc \au{Teaca, B.}, \au{Gorbunov, E.~A.}, \au{Told, D.}, \au{Bañón~Navarro,
  A.} \& \au{Jenko, F.}} \yr{2021}  \at{Sub-grid-scale effects in magnetised
  plasma turbulence}.  \jt{Journal of Plasma Physics}  \bvol{87}~(2),
  \pg{905870209}.

\bibitem[Teaca {\em et~al.\/}(2017)Teaca, Jenko \& Told]{Teaca:2017p1989}
{\sc \au{Teaca, B.}, \au{Jenko, F.} \& \au{Told, D.}} \yr{2017}
  \at{Gyrokinetic turbulence: between idealized estimates and a detailed
  analysis of nonlinear energy transfers}.  \jt{New J. Phys.}  \bvol{19},
  \pg{045001}.

\bibitem[Teaca {\em et~al.\/}(2014)Teaca, Navarro \& Jenko]{Teaca:2014p1571}
{\sc \au{Teaca, B.}, \au{Navarro, A.~B.} \& \au{Jenko, F.}} \yr{2014}  \at{The
  energetic coupling of scales in gyrokinetic plasma turbulence}.  \jt{Phys.
  Plasmas}  \bvol{21},  \pg{072308}.

\bibitem[Teaca {\em et~al.\/}(2012)Teaca, Navarro, Jenko, Brunner \&
  Villard]{Teaca:2012p1415}
{\sc \au{Teaca, B.}, \au{Navarro, A.~B.}, \au{Jenko, F.}, \au{Brunner, S.} \&
  \au{Villard, L.}} \yr{2012}  \at{Locality and universality in gyrokinetic
  turbulence}.  \jt{Phys. Rev. Lett.}  \bvol{109},  \pg{235003}.

\bibitem[Teaca {\em et~al.\/}(2019)Teaca, Navarro, Told, G{\"o}rler, Plunk,
  Hatch \& Jenko]{Teaca:2019p2154}
{\sc \au{Teaca, B.}, \au{Navarro, A.~B.}, \au{Told, D.}, \au{G{\"o}rler, T.},
  \au{Plunk, G.}, \au{Hatch, D.~R.} \& \au{Jenko, F.}} \yr{2019}  \at{A look at
  phase space intermittency in magnetized plasma turbulence}.  \jt{The
  Astrophysical Journal}  \bvol{886},  \pg{65}.

\bibitem[Tenbarge \& Howes(2013)]{Tenbarge:2013p1730}
{\sc \au{Tenbarge, J.~M.} \& \au{Howes, G.~G.}} \yr{2013}  \at{Current sheets
  and collisionless damping in kinetic plasma turbulence}.  \jt{The
  Astrophysical Journal Letters}  \bvol{771},  \pg{L27}.

\bibitem[Told {\em et~al.\/}(2015)Told, Jenko, TenBarge, Howes \&
  Hammett]{Told:2015p1712}
{\sc \au{Told, D.}, \au{Jenko, F.}, \au{TenBarge, J.~M.}, \au{Howes, G.~G.} \&
  \au{Hammett, G.~W.}} \yr{2015}  \at{Multiscale nature of the dissipation
  range in gyrokinetic simulations of alfv{\'e}nic turbulence}.  \jt{Phys. Rev.
  Lett.}  \bvol{115},  \pg{025003}.

\bibitem[Vech {\em et~al.\/}(2020)Vech, Kasper, Klein, Huang, Stevens, Chen,
  Case, Korreck, Bale, Bowen, Whittlesey, Livi, Larson, Malaspina, Pulupa,
  Bonnell, Harvey, Goetz, de~Wit \& MacDowall]{Vech_2020}
{\sc \au{Vech, D.}, \au{Kasper, J.~C.}, \au{Klein, K.~G.}, \au{Huang, J.},
  \au{Stevens, M.~L.}, \au{Chen, C. H.~K.}, \au{Case, A.~W.}, \au{Korreck, K.},
  \au{Bale, S.~D.}, \au{Bowen, T.~A.}, \au{Whittlesey, P.~L.}, \au{Livi, R.},
  \au{Larson, D.~E.}, \au{Malaspina, D.}, \au{Pulupa, M.}, \au{Bonnell, J.},
  \au{Harvey, P.}, \au{Goetz, K.}, \au{de~Wit, T.~D.} \& \au{MacDowall, R.}}
  \yr{2020}  \at{Kinetic-scale spectral features of cross helicity and residual
  energy in the inner heliosphere}.  \jt{The Astrophysical Journal Supplement
  Series}  \bvol{246}~(2),  \pg{52}.

\bibitem[Wan {\em et~al.\/}(2012)Wan, Matthaeus, Karimabadi, Roytershteyn,
  Shay, Wu, Daughton, Loring \& Chapman]{Wan:2012p1837}
{\sc \au{Wan, M.}, \au{Matthaeus, W.~H.}, \au{Karimabadi, H.},
  \au{Roytershteyn, V.}, \au{Shay, M.}, \au{Wu, P.}, \au{Daughton, W.},
  \au{Loring, B.} \& \au{Chapman, S.~C.}} \yr{2012}  \at{Intermittent
  dissipation at kinetic scales in collisionless plasma turbulence}.  \jt{Phys.
  Rev. Lett.}  \bvol{109},  \pg{195001}.

\bibitem[Weidl {\em et~al.\/}(2015)Weidl, Jenko, Teaca \&
  Schlickeiser]{Weidl:2015p1676}
{\sc \au{Weidl, M.~S.}, \au{Jenko, F.}, \au{Teaca, B.} \& \au{Schlickeiser,
  R.}} \yr{2015}  \at{Cosmic-ray pitch-angle scattering in imbalanced mhd
  turbulence simulations}.  \jt{The Astrophysical Journal}  \bvol{811},
  \pg{8}.

\bibitem[Zhao {\em et~al.\/}(2020)Zhao, Zank, Adhikari, Nakanotani, Telloni \&
  Carbone]{Zhao_2020}
{\sc \au{Zhao, L.-L.}, \au{Zank, G.~P.}, \au{Adhikari, L.}, \au{Nakanotani,
  M.}, \au{Telloni, D.} \& \au{Carbone, F.}} \yr{2020}  \at{Spectral features
  in field-aligned solar wind turbulence from parker solar probe observations}.
   \jt{The Astrophysical Journal}  \bvol{898}~(2),  \pg{113}.

\bibitem[Zhdankin {\em et~al.\/}(2012)Zhdankin, Boldyrev, Mason \&
  Perez]{Zhdankin:2012p1824}
{\sc \au{Zhdankin, V.}, \au{Boldyrev, S.}, \au{Mason, J.} \& \au{Perez, J.~C.}}
  \yr{2012}  \at{Magnetic discontinuities in magnetohydrodynamic turbulence and
  in the solar wind}.  \jt{Phys. Rev. Lett.}  \bvol{108},  \pg{175004}.

\bibitem[Zhou {\em et~al.\/}(2004)Zhou, Matthaeus \& Dmitruk]{Zhou:2004p21}
{\sc \au{Zhou, Y.}, \au{Matthaeus, W.~H.} \& \au{Dmitruk, P.}} \yr{2004}
  \at{Colloquium: Magnetohydrodynamic turbulence and time scales in
  astrophysical and space plasmas}.  \jt{Rev. Mod. Phys.}  \bvol{76},
  \pg{1015}.

\bibitem[Zocco \& Schekochihin(2011)]{zocco2011}
{\sc \au{Zocco, A.} \& \au{Schekochihin, A.~A.}} \yr{2011}  \at{Reduced
  fluid-kinetic equations for low-frequency dynamics, magnetic reconnection,
  and electron heating in low-beta plasmas}.  \jt{Physics of Plasmas}
  \bvol{18}~(10),  \pg{102309}.

\end{thebibliography}

\end{document}